\documentclass[aps,reprint,pra,twocolumn,amsmath,amssymb,aps,floatfix,nofootinbib]{revtex4-1}

\usepackage{hyperref,graphicx,color}
\usepackage{color,graphicx}
\usepackage[utf8]{inputenc}
\usepackage[T1]{fontenc}
\usepackage[english]{babel}
\usepackage{dcolumn,bm}
\usepackage{subfigure}
\usepackage{tikz}
\usepackage{calc}
\usepackage[bottom]{footmisc}
\usetikzlibrary{fit,matrix}
\hypersetup{colorlinks,%
    hypertexnames=true,%
    pdfauthor={Andre Brandst\"otter, Adrian Girschik, Philipp Ambichl, Stefan Rotter},%
    pdftitle={Control of Branched Flow in Optics},%
    pdfview=FitH,%
    pdfstartview=FitH,%
    pdfpagemode=UseNone,%
    bookmarksopen=true,%
    bookmarksnumbered=true,%
}

\newcommand\blfootnote[1]{%
  \begingroup
  \renewcommand\thefootnote{}\footnote{#1}%
  \addtocounter{footnote}{-1}%
  \endgroup
}

\begin{document}
\setlength{\tabcolsep}{1pt}
\title{Shaping the Branched Flow of Light through Disordered Media}
\author{Andre Brandst\"otter*}
\affiliation{Institute for Theoretical Physics, Vienna University of Technology (TU Wien), 1040, Vienna, Austria, EU}
\author{Adrian Girschik*}
\affiliation{Institute for Theoretical Physics, Vienna University of Technology (TU Wien), 1040, Vienna, Austria, EU}
\author{Philipp Ambichl}
\affiliation{Institute for Theoretical Physics, Vienna University of Technology (TU Wien), 1040, Vienna, Austria, EU}
\author{Stefan Rotter}
\affiliation{Institute for Theoretical Physics, Vienna University of Technology (TU Wien), 1040, Vienna, Austria, EU}

\begin{abstract}
Electronic matter waves traveling through the weak and smoothly varying disorder potential of a semi-conductor show branching behavior instead of a smooth spreading of flow. By transferring this phenomenon to optics, we show how the branched flow of light can be controlled to propagate along a single branch rather than through many of them at the same time. Our method is based on shaping the incoming wavefront and only requires partial knowledge of the system's transmission matrix. We show that the light flowing along a single branch has a broadband frequency stability such that we can even steer pulses along selected branches - a prospect with many interesting possibilities for wave control in disordered environments.
\end{abstract}

\maketitle
\blfootnote{*These authors contributed equally to this work.}
When waves propagate through a disorder landscape that is sufficiently weak and spatially correlated, they form branched transport channels in which the waves' intensity is strongly enhanced. This phenomenon of ``branched flow'' was first discovered for electrons gliding through semiconductor heterostructures \cite{Topinka_Coherent_2001}. Instead of an isotropic spreading into all possible directions, the electron density injected through a quantum point contact was observed to form esthetically very appealing branch patterns. This intriguing behavior can be attributed to the smooth background potential that is always present in such structures \cite{Jura_Unexpected_2007}, which acts like an array of imperfect lenses giving rise to caustics \cite{Kaplan_Statistics_2002} and thereby to distinct intensity enhancements along branches \cite{Heller_Thermal_2005, Aidala_Imaging_2007, Maryenko_How_2012}. 
Although first discovered as a nano-scale wave effect, branched flow was soon understood to occur on a wide range of length scales up to the formation of hot spots in tsunami waves as a result of the propagation through the rough ocean sea bed \cite{Heller_Refraction_2008,Hohmann_Freak_2010,Ying_Linear_2011,Metzger_Statistics_2014}.

While a number of previous studies already focused on the statistics of this phenomenon \cite{Kaplan_Statistics_2002,Metzger_Universal_2010,Metzger_Statistics_2014} and on its origins
\cite{Kaplan_Statistics_2002,Jura_Unexpected_2007,Ni_Origin_2011,Ni_Emergence_2012,Liu_Stability_2013,Liu_Classical_2015}, the question of how branched flow can be controlled and thereby put to use for steering waves through a complex medium has not been addressed so far. This is probably due to the fact that the possibilities to shape and manipulate electrons or ocean waves are, indeed, very limited. In other words, for the experiments where branched flow was observed so far, the incoming wavefront as well as the potential that the wave explores were considered as predetermined and immutable. These limitations are currently about to be overcome in a new generation of experiments, where coherent laser light was observed to exhibit branching when propagating through very thin disordered materials such as the surface layer of a soap bubble \cite{patsyk_interaction_2018}. Specifically, we expect that the transfer of branched flow to the optical domain will open up the whole arsenal of photonics to shape the wavefront of such branched light beams \cite{Mosk_Controlling_2012, rotter_light_2017}.

\begin{figure}[h!]
\includegraphics[clip,width=1.03\linewidth]{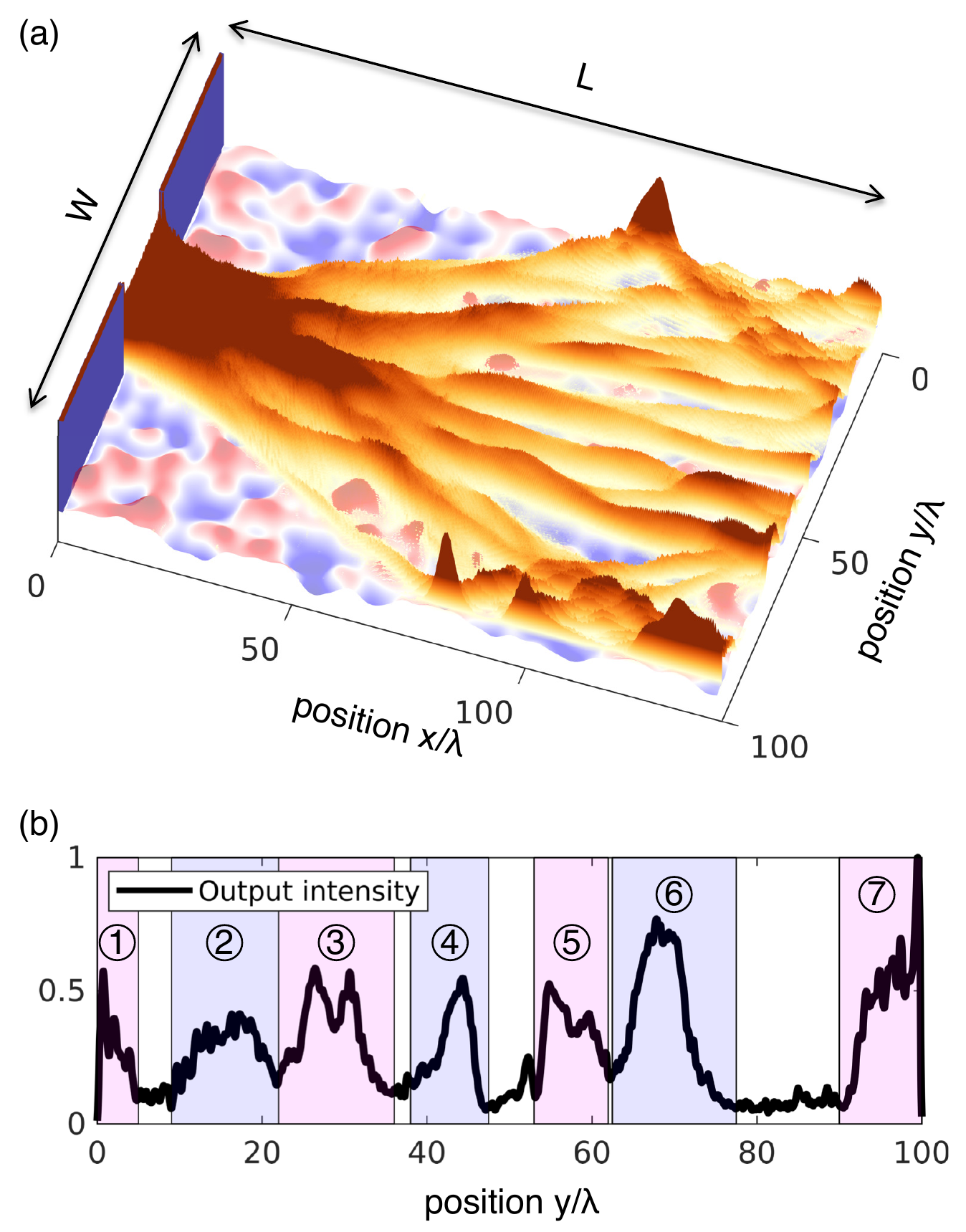}
\caption{(a) Illustration of the setup under study: a waveguide of width $W \approx 100 \lambda$ and length $L=1.4W \approx 140 \lambda$ is attached to an incoming lead on the left (through an aperture, blue) and an outgoing lead on the right. The smooth disorder potential in the waveguide is illustrated in blue/red colors. On top of the potential the wave intensity is plotted, corresponding to a superposition of the wave intensities of mode $1$ to $100$ (out of $200$ open modes) injected through the left lead, resulting in a pronounced branched strucuture. The main goal of our study is to separate these branches by suitably shaping the incoming wavefront in the left lead. (b) Intensity output profile as a function of transverse coordinate $y$ at $x=L$. The seven maxima labeled from 1 to 7 (highlighted in light blue/magenta) emerge due to the arrival of the branches at the output.}\label{fig:geometry}
\end{figure}

A particularly exciting question that we will explore here from a theoretical point of view will be whether optical wavefront shaping tools like spatial light modulators (SLMs) can be used to manipulate an incoming light beam in such a way that it follows only a single branch through a disorder landscape rather than many of them in parallel. The protocol that we will introduce based on our analysis will open up new ways of sneaking a beam of light through a disordered medium while maintaining its focus throughout the entire propagation distance - like a highway for light through a scattering medium.

The system we consider is shown in Fig.~\ref{fig:geometry}(a) and consists of a rectangular scattering region of length $L$ and width $W$ that is attached to two clean semi-infinite leads of the same width $W$ on the left and right (not shown). In transverse direction hard-wall boundary conditions are applied, i.e., the wavefunction is zero at these boundaries. In all the calculations reported below, we choose the number of propagating open lead modes to be $M=200$ and a fixed wavenumber $k = \mu \pi/W$ of the incoming light. For simplicity we set $W=\mu=200.01$ resulting in the following simple expressions for the wavenumber $k=\pi$ and the wavelength $\lambda=2$.
The length of the scattering region is chosen as $L=1.4 W \approx 140 \lambda$.

In analogy to the first observation of branched flow where electrons were injected through a constriction (quantum point contact) into a high-mobility electron gas \cite{Topinka_Coherent_2001}, we also include such a constriction in the form of an aperture of width $d=50.5 \approx 25\lambda$ between the left lead and the disordered scattering region. Whereas in many previous studies the width of the constriction was chosen such that it only allows for one or two modes to propagate through, the $50$ modes that we allow to pass feature many more tunable degrees of freedom as required for shaping the incoming wavefront. 
The smooth and long-range disorder necessary to observe branched flow is modeled by a spatially dependent index of refraction $n(\vec{r\,})$ throughout the whole scattering region indicated by the light red/blue color in Fig.~\ref{fig:geometry}(a). 
This disorder potential is characterized by its maximal refractive index $n_{\rm{max}}=1.1$, the minimal index at the vacuum value $n_{\rm{min}}=1$, and a finite spatial correlation length $\xi$, defined as the standard deviation of a Gaussian auto-correlation function 

\begin{align}
C(|\vec{r}-\vec{r}^{\,\prime}|) &= 
\langle (n^2(\vec{r}\,)-n^2_{\rm{min}})\cdot (n^2(\vec{r}^{\,\prime})-n^2_{\rm{min}})\rangle \\ \notag
& \propto 
\exp \left({\frac{-|\vec{r}-\vec{r}^{\,\prime}|}{2\xi}}\right) .
\end{align}
We choose $\xi=6$, which is three times longer than the operating wavelength $\lambda=2$.
The scalar scattering problem in this two-dimensional setup is described by the two-dimensional Helmholtz-equation
\begin{equation} \label{eq:helmholtz}
    [\Delta + k^2 n^2(\vec{r\,})]\psi(\vec{r\,}) = 0\mbox{,}
\end{equation}
with $\psi(\vec{r\,})$ representing the out-of-plane $z$-component of the electric field and $k=\omega / c$ being the incoming wavenumber. Employing the modular recursive Green's function technique \cite{Rotter_Modular_2000, Libisch_Coherent_2012}, we can efficiently evaluate the scattering states $\psi(\vec{r\,})$ and the unitary scattering matrix,
\begin{equation} \label{eq:scattering_matrix}
S= \left( \begin{array}{cc} r & t^{\prime} \\ t & r^{\prime} \end{array} \right)\mbox{,}
\end{equation}
with the transmission (reflection) matrix $t$ ($r$) containing the complex amplitudes $t_{ab}$ ($r_{ab}$) for transmission (reflection) from mode $b$ from the left lead to mode $a$ in the right (left) lead. 
The primed quantities $t^{\prime}$ and $r^{\prime}$ are the corresponding matrices for injection from the right.

In order to observe the branched flow of light, we inject the lead modes into the structure from the left and superimpose the corresponding wave intensities they give rise to. In the superposition we consider only the first $100$ lead modes to avoid high angle scattering and to ensure a high visibility of the individual branches. The branched structure in the propagation of waves through our setup is clearly visible in Fig.~\ref{fig:geometry}(a).

The challenge we rise to in a next step is to address these branches individually through a suitable coherent superposition of incoming modes in the left lead. The methods we choose for this purpose involve only the transmission matrix $t$ from Eq.~(\ref{eq:scattering_matrix}), which is available in optics through interferometric measurements involving an SLM \cite{Popoff_Measuring_2010, rotter_light_2017}. As the branched flow in our system naturally leads to a concentration of intensity at certain spots at the output, we find here that the knowledge of the transmission matrix $t$ for modes concentrated around these spots is sufficient for a clean separation of branches. In other words, we may restrict ourselves to those regions in space at the output, where the branches arrive. These regions are determined from the intensity profile at the output facet of our system at $x=L$, see Fig.~\ref{fig:geometry}(b). Seven intensity maxima corresponding to the arrival of different branches are clearly visible in Fig.~\ref{fig:geometry}(b) and highlighted in light blue/magenta. For each intensity maximum we manually set lower and upper boundaries, which are indicated by vertical lines in Fig.~\ref{fig:geometry}(b) and define a transmission matrix $\bar{t}$ connecting the incoming lead with the corresponding region at the output (labeled from 1 to 7). The elements $\bar{t}_{ab}$ of this matrix hence describe the coherent transmission amplitudes from all points $b$ at the input (we choose 200 equidistant points in the input lead corresponding to 200 open lead modes) to all points $a$ around a specific intensity maximum at the output.

\begin{figure*}
\includegraphics[clip,width=\textwidth]{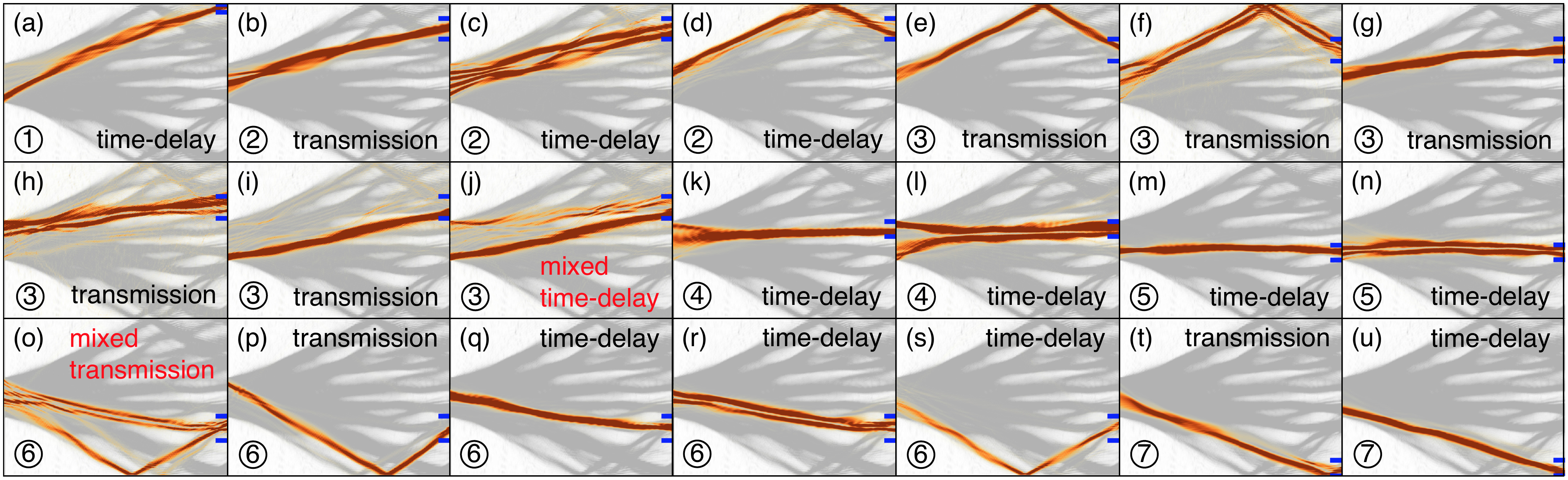}
\caption{Transmission and time-delay eigenstates of the system shown in Fig.~\ref{fig:geometry} calculated from the transmission matrices $\bar{t}$ connecting the incoming lead with the seven different output regions marked in Fig.~\ref{fig:geometry}(b) (see blue horizontal lines here). The branched structure [see Fig.~\ref{fig:geometry}(a)] is drawn as a gray background. We show all eigenstates that are selected by our procedure [steps (i)-(iii) described on page 4] with black text labels and two examples of eigenstates that are filtered out by our procedure since they address two different branches at once (red label). The states are ordered according to their output region on the right (see numbers 1 to 7). The average transmittance of all states with black label is around $89\%$ into the region marked by the blue lines.}\label{fig:branches}
\end{figure*} 

Our first approach to achieve clean branch separation is to employ a singular value decomposition (SVD) of $\bar{t}$, i.e.,
\begin{equation} \label{eq:svd}
    \bar{t}=U \Sigma V^{\dagger},
\end{equation}
allowing us to access the transmission eigenvalues $\tau_i$ in this truncated spatial basis as contained in the real diagonal matrix 
$\Sigma=\rm{diag}\,(\sqrt{\tau_i})$. The matrix $U$ consists of eigenvectors of $\bar{t}\bar{t}^{\,\dagger}$ and $V$ consists of eigenvectors of $\bar{t}^{\,\dagger}\bar{t}$. The largest transmission eigenvalues $\tau_i$ correspond to those transmission eigenchannels $\vec{\tau}_i$ (contained in the columns of $V$) that transmit the most intensity to the desired region at the output. At this point one might be tempted to think that these highly transmitting channels will already constitute the branches we are after. To test this hypothesis, we inject for each of the seven transmission matrices $\bar{t}$, the corresponding transmission eigenchannels with the largest transmission eigenvalues. Checking the corresponding scattering wavefunctions [see Fig.~\ref{fig:branches}(b),(e)-(i),(p),(t)], we find that a clean branch separation is, indeed, possible for a number of cases. We also find, however, that several among the highly transmitting eigenchannels follow two different branches in parallel instead of only one. Figure \ref{fig:branches}(o) shows an example of such a state, where one can clearly see a mixing of one branch propagating directly into the selected region (marked with blue lines at the output) with another branch bouncing off the lower boundary. Demanding high transmission into a desired region by choosing high transmission eigenchannels is thus clearly not enough to guarantee clean branch separation since high transmission can also be obtained by propagating along multiple branches at once. In a possible optical experiment such a mixing can be expected to be even more prevalent than in our numerical example, simply because in such an experiment many more branches are typically available.

In order to be able to address also such mixed branches individually, we now introduce a more efficient method. Specifically, our aim is to set up an approach in terms of the scattering matrix $S$ and the Wigner-Smith time-delay operator \cite{Wigner_Lower_1955, Smith_Lifetime_1960, xiong_spatiotemporal_2016} derived from it,
\begin{equation} \label{eq:Q}
    Q= -i S^{-1} \frac{\rm{d}S}{\rm{d}\omega} \mbox{.}
\end{equation}
Eigenstates of $Q$, also known as principal modes, are associated with scattering states that have a well-defined delay time and the remarkable property that their output profile is very robust to frequency changes \cite{fan_principal_2005, rotter_generating_2011, carpenter_observation_2015, gerardin_particlelike_2016, ambichl_super-_2017, bohm_situ_2018}. Some of these eigenstates have the additional feature of having a particle-like wavefunction, i.e., the scattering states follow classical particle trajectories \cite{rotter_generating_2011, gerardin_particlelike_2016, bohm_situ_2018}. Modifying the Wigner-Smith time-delay operator for the present purpose now allows us to separate those eigenstates of $\bar{t}^{\,\dagger}\bar{t}$ with the largest transmission eigenvalues $\tau_i$ by their time-delay \cite{ambichl_focusing_2017, bohm_situ_2018}. The key idea here is that two branches that may both be highly transmitting [like those in Fig.~\ref{fig:branches}(o)] can be distinguished by their different time-delays (as determined by the different branch lengths). To be specific, we only work with those $N$ transmission eigenvalues $\tau_i$ that are larger than some value $\eta$ and derive 
the matrices $u$, $v$ and $\sigma$ from $U$, $V$ and $\Sigma$ by truncating all rows and columns corresponding to $\tau_i < \eta$.
With these truncated matrices we can now replace the terms in Eq.~(\ref{eq:Q}), 
\begin{equation}
\frac{\rm{d}S}{\rm{d}\omega} \to u u^{\dagger} \frac{{\rm{d}} \bar{t}}{{\rm{d}}\omega} v v^{\dagger} \quad \rm{and} \quad S^{-1} \to \rm{''}\bar{t}^{-1}\rm{''} \to v \sigma^{-1} u^{\dagger},
\end{equation}
to arrive at the reduced time-delay operator $q$,
\begin{equation} \label{eq:q}
    q = -i v {\sigma}^{-1} u^{\dagger}u u^{\dagger} \frac{{\rm{d}} \bar{t} }{{\rm{d}}\omega}v v^{\dagger},
\end{equation}
that operates in the sub-space of highly transmitting states only. Note that Eq.~(\ref{eq:q}) involves a quasi-inverse ''$\bar{t}^{-1}$'' of the rectangular matrix $\bar{t}$ whose regularity is guaranteed by the restriction to only those transmission eigenvalues $\tau_i$ that are larger than  the cut-off value $\eta$. In practice, a value of $\eta=0.8$ proved suitable for all our calculations. Note that, due to the non-unitarity of $\bar{t}$, the eigenvalues of the reduced time-delay operator $q$ in Eq.~(\ref{eq:q}) are complex [in contrast to the real eigenvalues of the Wigner-Smith time-delay operator $Q$ in Eq.~(\ref{eq:Q})]. The imaginary parts of the complex eigenvalues are, however, very small and the real parts can still be used as a good measure for the physical delay times \cite{bohm_situ_2018}.

To put this method directly to the test, we turn our attention to the state shown in Fig.~\ref{fig:branches}(o) featuring a mixture of two branches with different path lengths and, correspondingly, different time-delays. A singular value decomposition of $\bar{t}$ reveals that it contains $9$ singular values larger than $\eta=0.8$. With this knowledge we can now construct $q$ according to Eq.~(\ref{eq:q}) and, indeed, find among its eigenstates the desired wave fields that follow the two involved branches individually [see Fig.~\ref{fig:branches}(r) and (s)].

Restricting the construction of time-delay eigenstates to the sub-space of high transmission thus yields already very good results. Using this method, we, however, also observed a few time-delay eigenstates that mix two different branches as, e.g., shown in Fig.~\ref{fig:branches}(j). These two branches, on the other hand, turn out to be individually addressable through those transmission eigenstates $\vec{\tau}_i$ with the smallest time-delays [see Fig.~\ref{fig:branches}(g) and (i)] (the time-delay of a transmission eigenstate $\vec{\tau}_i$ can simply be calculated by taking the expectation value with the time-delay operator $q$, i.e., $\vec{\tau}_i^{\dagger}q\vec{\tau}_i$).

\begin{figure}[t]
\includegraphics[clip,width=1\linewidth]{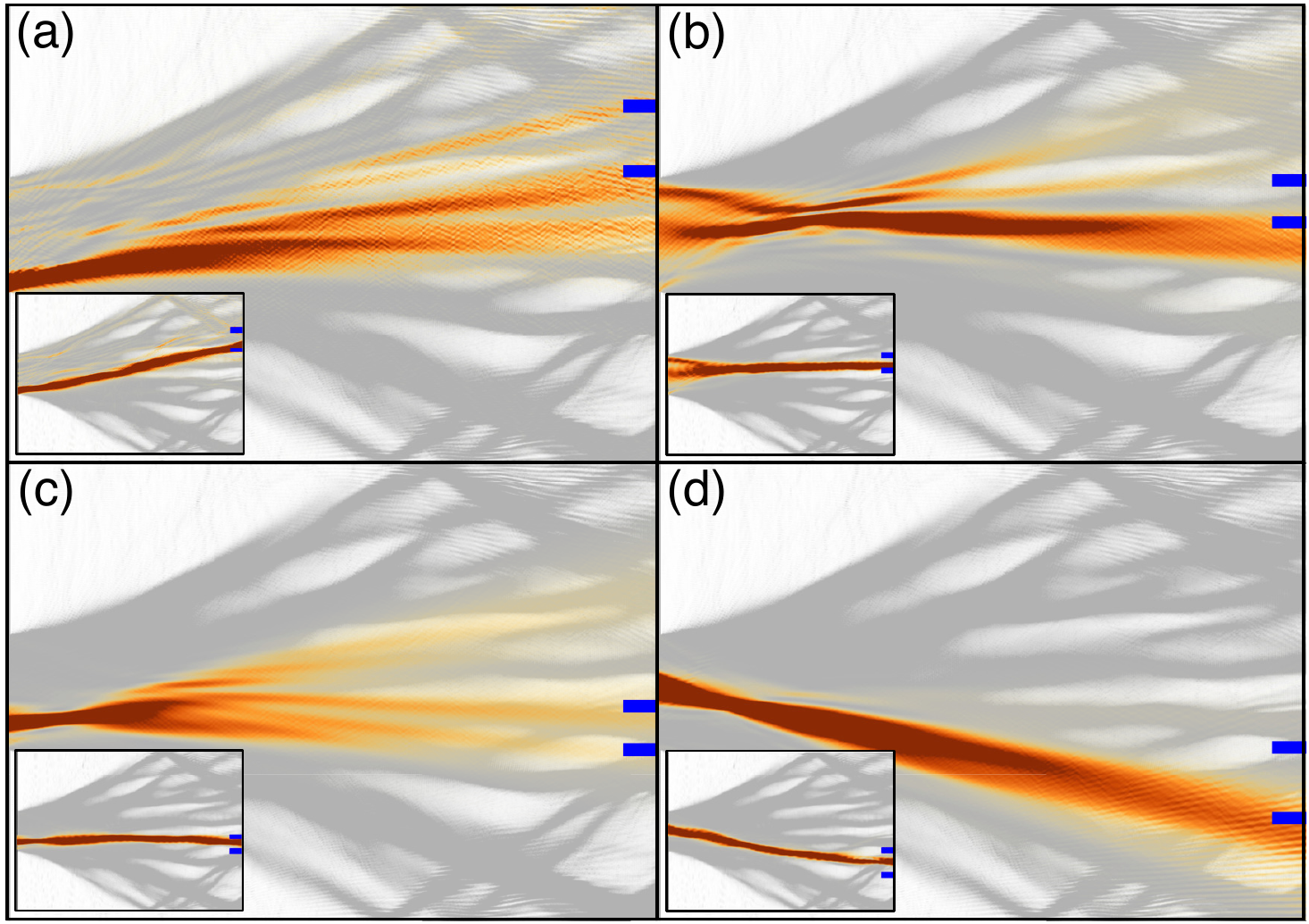}
\caption{(a)-(d) Eigenstates from Fig.~\ref{fig:branches}(i),(k),(m),(q) injected into an empty waveguide without the disordered refractive index. The fact that removing the disorder potential leads to a defocussing demonstrates that the formation of collimated branch states crucially relies on the presence of the underlying disorder landscape. The blue lines indicate the region where the branch exits the scattering region in the presence of the disorder (see insets).}\label{fig:branches_empty}
\end{figure} 
One may thus also decide to turn the above strategy on its head and look for transmission eigenstates in the sub-space of short time-delays. Since neither one of these opposite strategies seems to have an a-priori advantage, we now combine them with each other in a synergistic way to improve our results even further: (i) We evaluate all eigenstates $\vec{\tau}_i$ of $\bar{t}^{\,\dagger}\bar{t}$ and $\vec{q}_i$ of $q$, for all of the seven transmission matrices $\bar{t}$ corresponding to regions of maximum intensity shown in Fig.~\ref{fig:geometry}(b). (ii) We select those states that are identical in both eigenstate sets, since they turn out to be individual branch excitations in all of the observed cases. To do this, we project each eigenvector $\vec{q}_i$ onto each eigenvector $\vec{\tau}_i$, such that we end up with the matrix elements $m_{ij}= \vec{q}_i^{\,\dagger} \vec{\tau}_j$. (The matrix $m$ is not unitary since the eigenstate-sets are not complete). For the case that two eigenvectors are the same, the matrix $m$ has only one significant non-zero element in the corresponding row/column. Practically, we consider two eigenvectors to be the same when $|m_{ij}|>0.9$. (iii) In a last step, we deal with those eigenstates that consist of more than one contribution from the respective other eigenstate set, i.e., that have more than one non-zero element in the corresponding row/column of $m$. Our task here is to select those states that consist of only single branches and to discard those states that propagate along more than one branch at once. Since, however, the coefficients $m_{ij}$ do not indicate per se which states consist of single branches only, we first need to translate the eigenvector coefficients to a corresponding angular profile at the input aperture. Checking, in a next step, if this angular input pattern is sufficiently collimated provides us finally with the desired indicator for the excitation of a single branch (see Appendix for details).

Following the above three steps (i)-(iii), which notably rely only on the experimentally accessible transmission matrices $\bar{t}$, we obtain well-separated branch states (see all states in Fig.~\ref{fig:branches} with a black text label) that stay collimated throughout the entire scattering region and that feature an average transmittance of over $89\%$ into the small output region. These results show that our method leads to a channeling of waves through the disordered region and to a well-controlled branched flow. An interesting detail that we emphasize here is that our approach not only yields a single state for each individual branch but, in fact, also states that propagate along the same branch but with a higher transverse quantization; see, e.g., Fig.~\ref{fig:branches}(c),(f),(h),(l),(n),(r) \cite{rotter_generating_2011}.

\begin{figure}[t]
\includegraphics[clip,width=1\linewidth]{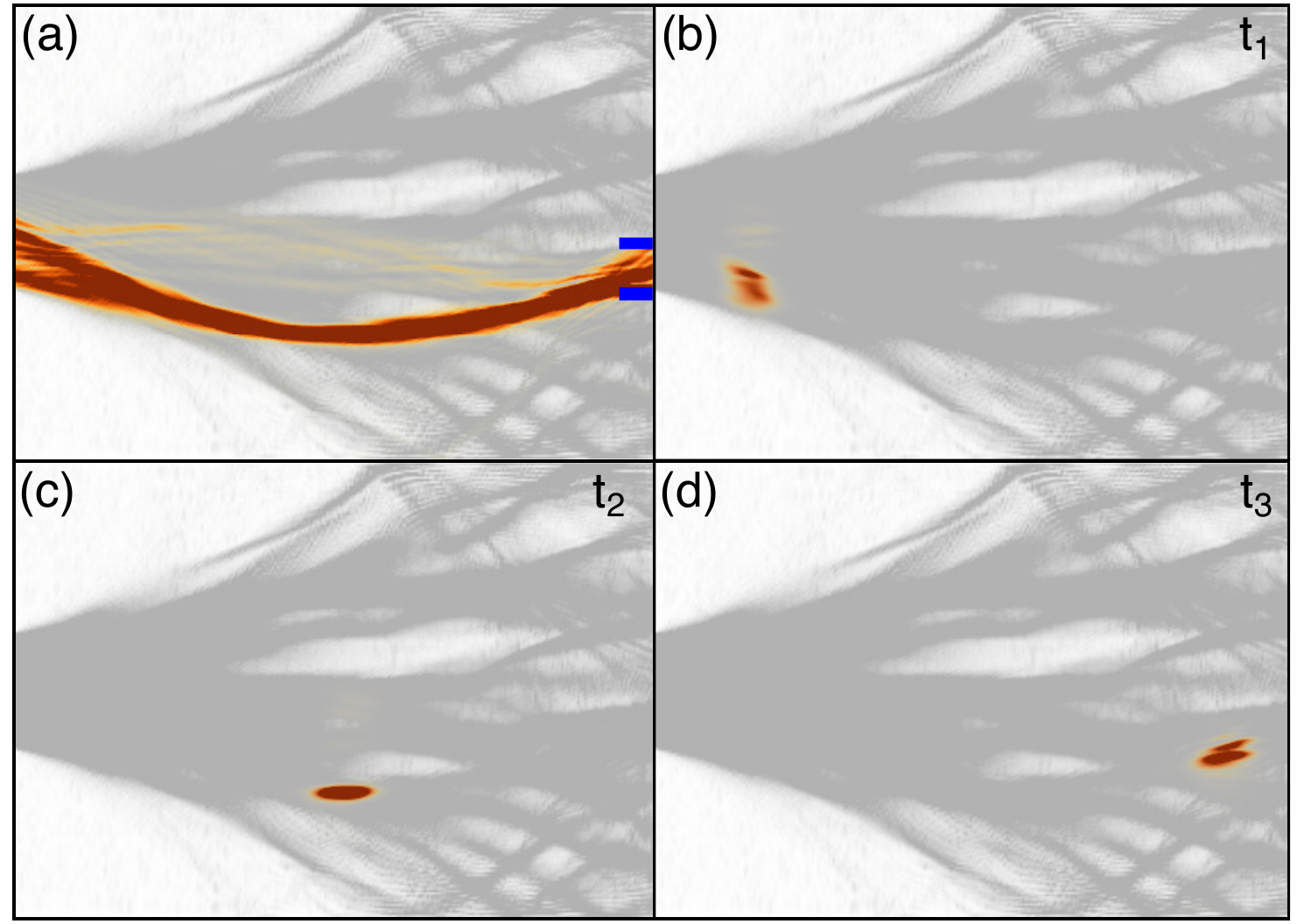}
\caption{(a) Time-delay eigenstate propagating along one single branch in a disorder landscape [to prove the general applicability of our approach, a different disorder realization was used as in Fig.~\ref{fig:geometry}(a)]. (b)-(d) Pulse propagating along the branch shown in (a) at three different time-steps ($t_1<t_2<t_3$). The pulse remains spatially confined while traversing the disorder along the branch shown in (a). The Fourier spectrum of the pulse is Gaussian-shaped with a standard deviation of $\sigma \approx 0.034 k$, with $k$ being the wavenumber.}\label{fig:pulse}
\end{figure} 

To underscore the non-trivial nature of these collimated branch states that we identify here for the first time, we inject several of the states shown in  Fig.~\ref{fig:branches} into a clean cavity without any disorder. The results are displayed in Fig.~\ref{fig:branches_empty}, showing that these states feature a considerably reduced collimation as compared to the case including the disorder (see figure insets). This observation demonstrates that the states we identify here do not just rely on a trivial injection with a narrow angular distribution at the input and that the disorder plays a crucial role for the states' collimation.

In a last part of this study, we also demonstrate explicitly that our collimated single-branch states can be sufficiently stable in frequency to allow for the transmission of pulses along a branch. Consider here, as an example, the time-delay eigenstate shown in Fig.~\ref{fig:pulse}(a) that propagates along a certain branch. Taking a superposition of this branch state at different frequencies to form a Gaussian wave packet, we obtain the pulse propagating along the selected branch as shown in Fig.~\ref{fig:pulse}(b)-(d) at three different time-steps ($t_1<t_2<t_3$). We observe that the pulse transits the system while staying on the selected branch throughout the entire transmission process.

In summary, this work demonstrates how to control the flow of waves through a correlated and weak disorder potential landscape. Such systems give rise to branches along which incoming waves travel through the disorder. We introduce a method that allows us to inject waves in such a way that almost all the flow travels along a single branch alone. This non-trivial finding can even be extended to the temporal domain, as we show by creating pulses that remain on a single branch throughout the entire transmission process. Implementing such concepts in optics requires only a small sub-part of the transmission matrix and is thus within reach of present-day technology. 

We thank Miguel A.~Bandres for fruitful discussions on his recent experiments \cite{patsyk_interaction_2018}. We acknowledge support by the European Commission under project NHQWAVE (MSCA-RISE 691209). AB is a recipient of a DOC Fellowship of the Austrian Academy of Sciences at the Institute of Theoretical Physics of Vienna University of Technology (TU Wien). The computational results presented have been achieved using the Vienna Scientific Cluster (VSC).

\setcounter{secnumdepth}{-1}
\begin{appendix}
\section{Appendix: Spatial and angular profile of eigenstates  \label{App}}

In order to find individual branch excitations among all eigenstates $\vec{q}_i$ and $\vec{\tau}_i$, it is essential to determine if either a time-delay eigenstate $\vec{q}_i$ or a transmission eigenstate $\vec{\tau}_i$ addresses only one single branch rather than many at the same time. As we show here, the spatial and/or angular distribution of an eigenstate at the input aperture provides us with sufficient information to perform this task since an eigenstate exciting only one branch will be spatially more confined and will radiate into a smaller angular region than a state addressing multiple branches.  Assuming that the transmission matrix $\bar{t}$ is measured in the spatial pixel basis, the eigenvectors $\vec{q}_i$ and $\vec{\tau}_i$ are naturally given in this spatial basis as well. By plotting the absolute value of the coefficients $|c_n^y|$, where $n$ is the $n$-th component of the vector $\vec{q}_i$ or $\vec{\tau}_i$, as a function of the transverse coordinate at the aperture ($x=0$), we can easily generate the spatial distribution of an eigenstate.  

To estimate the angular distribution of an eigenstate at the aperture, we work with the Hermitian operator $k_y = -id/dy$ measuring the transverse $y$-component of the wavevector. The eigenvalue equation of the $i$-th eigenvector $\vec{k}^{(i)}_y$ of this operator reads:
\begin{equation}
k_y \vec{k}_y^{(i)}= \lambda^{(i)} \vec{k}_y^{(i)},
\end{equation}
where $\lambda^{(i)}$ is the $i$-th eigenvalue. Since a well-defined transverse wavevector component corresponds to a well-defined angle of incidence, we can now decompose the eigenvectors $\vec{q}_i$ and $\vec{\tau}_i$ into the momentum basis spanned by the vectors $\vec{k}_y^{(i)}$ and analyze the different angular components $|c_n^k|$.

\begin{figure}[t!]
\includegraphics[clip,width=0.9\linewidth]{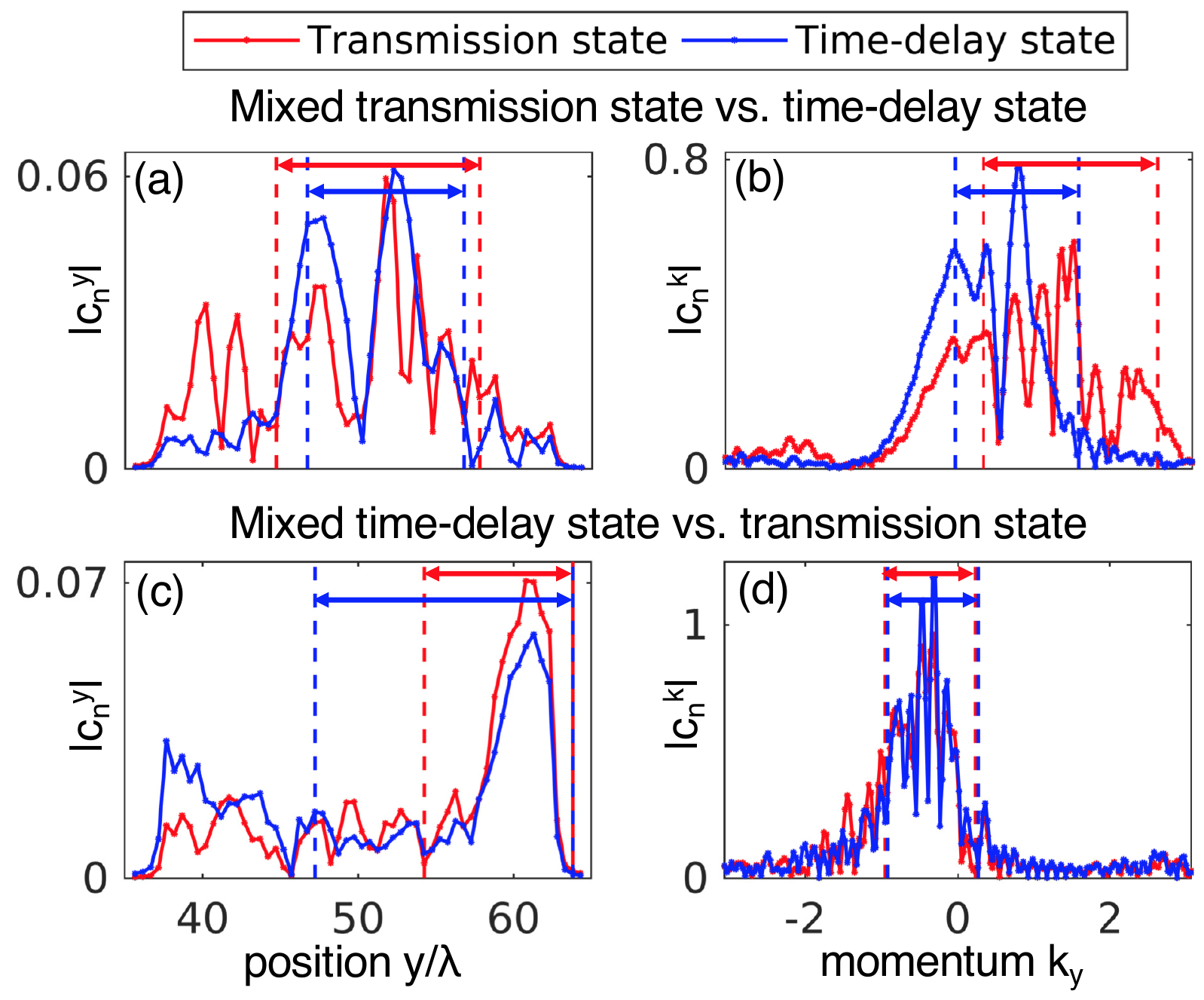}
\caption{(a),(b) Spatial (left) and angular distribution (right) at the input aperture (located between $y/\lambda \approx 37$ and $y/\lambda \approx 62$) of the (mixed) transmission (red) and time-delay state (blue) shown in Fig.~\ref{fig:branches}(o) and Fig.~\ref{fig:branches}(r) of the manuscript, respectively. The different widths of the distributions indicate that the transmission state is very likely to excite more than one branch at once. (c),(d) Spatial (left) and angular distribution (right) of the (mixed) time-delay (blue) and transmission (red) state shown in Fig.~\ref{fig:branches}(j) and Fig.~\ref{fig:branches}(i). From the widths of the spatial distributions shown in (c) we can conclude that the transmission state is more likely to excite only one single branch, which is confirmed by the wave plots in the manuscript. The widths of the normalized distributions are quantified by the interval around the maximum value of the distribution (indicated by the vertical dashed lines) in which 60\% of the distribution lies.}\label{fig:separation}
\end{figure} 

Figure \ref{fig:separation}(a) and (b) display the spatial and angular components of the transmission eigenstate shown in Fig.~\ref{fig:branches}(o) (red) and the time-delay eigenstate in Fig.~\ref{fig:branches}(r) (blue). We see that the spatial profile of the transmission state is broader and that it features more angular components than the time-delay state. We can therefore conclude that the transmission state is more likely to address multiple branches, whereas the time-delay state addresses only one single branch, which is confirmed by the wave plots shown Fig.~\ref{fig:branches}(o) and Fig.~\ref{fig:branches}(r). In Fig.~\ref{fig:separation}(c) and (d) we can see the same distributions for the time-delay state shown in Fig.~\ref{fig:branches}(j) and the transmission state shown in Fig.~\ref{fig:branches}(i). From Fig.~\ref{fig:separation}(c) we deduce that the time-delay state consists of more than one branch due to the larger spatial distribution, which is confirmed by the wave plots in the manuscript. We successfully applied this procedure to all eigenstates $\vec{q}_i$ and $\vec{\tau}_i$ from which we can conclude that the spatial and angular distribution of the time-delay and transmission states can be used to find those states out of both eigenstate sets (time-delay and transmission eigenstates) that excite only one single branch. 

\end{appendix}

\bibliographystyle{apsrev4-1}

\end{document}